# Investigation of robust population transfer using quadratically chirped laser interacting with two-level system

Fatemeh Ahmadinouri [1], Mehdi Hosseini[1*], Farrokh Sarreshtedari[2]

*[1]Department of physics, Shiraz University of Technology, Shiraz, 313-71555, Iran*

*[2]Magnetic Resonance Research Laboratory, Department of Physics, University of Tehran, Tehran 143-9955961, Iran*

We have proposed and demonstrated a fast and robust method of population transfer between two quantum states using a quadratically chirped laser source. Incorporating the Jaynes-Cummings in a full quantum description of the interaction, and numerically solving the time-dependent Schrödinger equation, transition probabilities have been obtained and the condition of the adiabatic passage is investigated. In this scheme, a laser source has been swept quadratically in time for arbitrarily engineering the transition probabilities. The results show that complete and robust population transfer could be selectively achieved by appropriate adjusting of the laser chirping parameter, the center frequency and the coupling strength which the time of the complete transition could be drastically decreased compared to linearly traditional chirped laser. Furthermore, another feature of using the quadratically chirped laser is the stimulation of intermediate transitions under the nonadiabatic passage.

---

* hosseini@sutech.ac.ir

# I. INTRODUCTION

Manipulation and control of quantum states are substantial issues in various fields of the atomic physics including quantum information [1-3], atomic clocks [4], spectroscopy [5], surface plasmon polariton [6,7], chemical reactions [8], Bose-Einstein condensate [9], collision dynamics [10], laser cooling [11], interferometry [12] and magnetic resonance [13]. Among these applications, much attention is paid to keep these quantum systems at the high fidelity and robustness [14-16].

In this regard, Landau-Zener technique is one of the important methods of the population transfer which is widely used in atomic and molecular physics. In this scheme, two energy levels under the influence of time-dependent Hamiltonians share an avoided crossing when the external parameters are swept linearly in time [17, 18]. Although this technique was introduced in 1932 [19, 20], today it is widely used in different experimental quantum systems like Landau-Zener in molecular nanomagnets [21], Bose–Einstein condensates [22] and Landau-Zener-Stuckelberg interferometry in superconducting circuits [23].

Another approach in population transfer is the incorporation of $\pi$ pulses that happen when the laser frequency is in resonance with the atomic energy levels and the flip angle is equal to $\pi$. It is worth mentioning that albeit in this method the fast and full population transfer is possible but it is significantly sensitive to any changes in the pulse area and inhomogeneities of the sample. Consequently, the robustness of this technique is quite low [24] Nevertheless, this population transfer approach is widely used in magnetic resonance [11].

Among the different popular strategies, the adiabatic passage is a population transfer method with high robustness [25, 26]. However, in order to assure the adiabatic condition, the transition is slowly occurred. One of the adiabatic passage methods for the three-level system is the Stimulated Raman Adiabatic Passage (SRIRAP). This method has emerged as a selective population transfer technique between two quantum states by an intermediate state in which under the adiabatic conditions two resonant fields are coupled to the three-level systems [27]. The population transfer through STIRAP approach and its different applications has been widely investigated and studied [1].

In addition, significant population transfer can be achieved using the chirped (or frequency-swept) pulse laser [28-36, 37]. Raman Chirped Adiabatic Passage (RCAP) is another method of population transfer based on adiabatic passage [38, 39]. In this layout, the laser frequency is drifted in time. If the frequency is chirped adequately slowly so the adiabatic condition remains, the transition appears in the region of an avoided crossing.

Several methods have been employed to achieve the fast population transfer in the adiabatic condition so that the system is preserved in high fidelity. Such methods are known as "shortcut to adiabaticity" [40, 41]. Among these techniques, counter-diabatic quantum driving [42, 43] and inverse engineering based on Lewis-Riesenfeld can be mentioned [44, 45].

Furthermore, in many works, the effects of perturbation on the population transfer of systems including dephasing and dissipation problems, environmental effects such as noise and temperature effect have been investigated [46-50].

The two-level approximation is a suitable model for the analytical and numerical description of many quantum systems. Many complex schemes and multi-level systems can also be solved by reducing them into two-level systems [46, 51].

The light-atom interaction is generally realized by semi-classical and full-quantum descriptions. Typically, in the population transfer methods mentioned above, the light-atom interaction is described semi-classically [6, 52]. The predominant model used to investigate the population transfer in full-quantum approach is the Jaynes-Cummings model. In this scheme, it is assumed that a quantized electromagnetic field (a single mode) interacts with a two-state atom [13].

Here, we have described a fast and robust population transfer in nonadiabatic passage regime in a two-level system by Jaynes-Cummings model. In this scheme, the interaction of the quantum system with the laser field is studied when laser frequency is swept (chirped) quadratically in time. The main advantage of this approach into the counter-diabatic technique simplifies the experimental implement of a quadratically smooth sweep of laser frequency compared to linear chirped accompanied by sudden changes in laser frequency [43]. In addition, the evolution of eigenstates and region of avoided crossing have been demonstrated in this research.

## II. THEORETICAL MODEL

In this work, the evolution of the energy levels and transition probabilities of a two-level quantum system have been investigated when a chirped laser source interacts with the atom. In this study, the Hamiltonian of the system includes the terms for the atom, laser and the interaction between them respectively as follow [13]:

$$H = -\frac{1}{2}\hbar\omega_0\sigma_z + \hbar\omega_L(a^\dagger a + \frac{1}{2}) + \hbar\omega_1(a\sigma_+ + a^\dagger\sigma_-)$$ (1)

Which is the well-known Jaynes-Cummings model. Where $\sigma_-$, $\sigma_+$, $\sigma_z$ are Pauli spin matrices. $a$, $a^\dagger$ are annihilation and creation operators respectively. $\hbar$ is Planck constant. Moreover, $\omega_0$ is the atomic resonance frequency, $\omega_L$ is the laser frequency and $\omega_1$ describes the coupling strength between atom-laser which is proportional to the atomic dipole moment and field strength [53].

By considering eigenstates |g, n+1⟩ and |e, n⟩ as ground state and excited state respectively where $n$ is the number of photons as well as acting the Hamiltonian on the system, eigenstates the final Hamiltonian can be achieved [54]. In these calculations, $n$ is considered one. It should be noted that the Jaynes-Cummings Hamiltonian is acquired by applying the rotating wave approximation (RWA). For validity of this approximation the system must be near resonance and the coupling strength ($\omega_l$) should be less than the atomic resonance frequency ($\omega_0$) [55].

Finally, by numerically solving the time-dependent Schrödinger equation by Runge–Kutta method, the results are obtained.

When the adiabatic evolution occurs, the system stays on an eigenstate of the Hamiltonian as the system parameters are changed. Mathematically, the adiabatic condition is provided when [33]:

$$\left| \frac{d\theta}{dt} \right| << \sqrt{\Omega^2 + \Delta\omega_L{}^2} \qquad (2)$$

Where $\Delta\omega_L$ is frequency detuning and mixing angle [53] is given by $\theta = \frac{1}{2}\tan^{-1}\frac{2\omega_\perp}{\Delta\omega_L}$ . $\Omega$ is Rabi frequency which is equal to $\Omega = \sqrt{\Delta\omega_L{}^2 + (2\omega_l)^2}$ [53]. Under these conditions and assuming $\Delta\omega_L << 2\omega_l$ the adiabatic approximation is achieved.

In this work, we acquire the impact of the quadratically chirped laser field parameters on the population transfer and the quantum control robustness as well as the condition of the adiabatic passage is investigated. For this purpose, we have measured mean values in the 10% Final Probability of Ground state which is called FPG. Furthermore, the evolution of the adiabatic transition probabilities is studied.

It is also worth mentioning that for simplicity the system parameters have been dimensionless and all the parameters have been normalized as follows:

$$\tilde{\omega}_0 = 1, \tilde{\omega}_1 = \frac{\omega_1}{\omega_0}, \tilde{\omega}_{L0} = \frac{\omega_{L0}}{\omega_0}, \tilde{t} = t\,\omega_0, \tilde{\alpha} = \frac{\alpha}{\omega_0^3}, \tilde{E} = \frac{E}{\hbar\omega_0} \qquad (3)$$

It should be noted that in all calculation, the time intervals are chosen somehow to ensure that the final probabilities have reached to the stable values. In addition, it is assumed that the system has been in the ground state for a long time ago and for guarantee of the RWA validity, the coupling strength is considered as $\tilde{\omega}_1 \prec 1 (\omega_1 \prec \omega_0)$.

### III. RESULTS AND DISCUSSION

Here, the laser frequency is considered as a square function of time:

$$\omega_L = \alpha t^2 + \omega_{L0} \qquad (4)$$

Where $\alpha$ is the chirping parameter, $\omega_{L0}$ is the center frequency and $t$ is the time.

In the quadratic chirped method, the adiabatic condition is obtained as:

$$|\alpha t| << (2\omega_1)^2 \qquad (5)$$

In this regard, by decreasing the chirping parameter and the transition time or by increasing the coupling strength the adiabatic condition can be satisfied. It should be also noted that in this relation, the transition time depends on the chirping parameter so that by increasing the chirping parameter, the transition time is reduced.

In Fig. 1 in order to the investigation of population transfer induced by the quadratically chirped laser, state probability is plotted for different laser field parameters (probability "1" means complete population transfer). In Fig. 1(a) transition probability versus chirping parameter and time for $\tilde{\omega}_{L0}$=0.1 and $\tilde{\omega}_1$=0.1 is depicted. In this diagram by increasing the chirping parameter, the final transition probability changes. It can be seen that some unstable transitions happen at the time around zero where by increasing the chirping parameter, the number of these peaks are reduced. Fig. 1(b) shows the dependence of the transition probability to the center frequency when $\tilde{\alpha}$=0.002 and $\tilde{\omega}_I$=0.1. In Fig. 1 (c, d) the results of transition probability is shown when the coupling strength is changed for $\tilde{\alpha}$=0.002 and $\tilde{\omega}_{L0}$=0.1. By increasing the coupling strength, the final probability is varied and the probability of intermediate transition is increased while the number of peaks are constant (Fig. 1(c)). By further increasing of the coupling strength, the final probability is reached zero but the intermediate transition probability is fixed and only the fluctuation of curves are added (d). It should be noted that such behavior is typical in the Landau-Zener model [56].

Fig. 2 shows transition probability versus time for linearly chirped laser source which $\tilde{\alpha}$=0.002, $\tilde{\omega}_I$=0.1 and $\tilde{\omega}_{L0}$=1. In this figure, it is evident that the transition time of quadratically chirped laser (Fig. 1) is drastically decreased compared to the form of linearly chirped laser.

In Fig. 3 eigenstates versus time are illustrated for the different parameter. In Fig. 3(a, b) energy diagram is depicted for the different chirping parameter and $\tilde{\omega}_I$=0.1, $\tilde{\omega}_{L0}$=0.1. This figure reveals that for very low chirping parameter two areas of avoided crossing exist. By increasing the chirping parameter, the areas of avoided crossing are reduced to one area. In Fig. 3 (a, c) evolution of the energy versus time is plotted for different center frequency while $\tilde{\alpha}$=0.002, $\tilde{\omega}_I$=0.1. The behaviors are similar to the behavior of chirping parameter variations. Evolution of energy versus time for different coupling strength is demonstrated in Fig. 3 (a, d) when $\tilde{\alpha}$=0.002, $\tilde{\omega}_{L0}$=0.1.

The impact of coupling strength on the evolution of eigenstates reveals that by increasing the coupling strength, the area of avoided crossing is increased and this behavior is in the agreement with increasing transition probability. In contrast, we find that by further increase of the coupling strength, this trend is reversed and transition probability is decreased (Fig. 1(c, d)) [56].

The FPG versus center frequency and coupling strength for different values of the chirping parameters are illustrated in Fig. 4 (the blue regions reveals the complete population transfer). This figure shows that for very low coupling strength, no population transfer is visible which it is corresponding to the fact that laser field strength is very low. By increasing the coupling strength, the transition probability is also increased as far as the FPG goes to 1 in certain coupling strength and center frequency, thus complete transition occurs. Henceforth, the FPG is decreased so that it is reached to zero which is in agreement with the results of Fig. 1 (c, d). According to the adiabatic approximation in the quadratic chirped method (Eq. 5), the transition occurs in the region of the nonadiabatic passage. In order to assure the adiabatic passage, the coupling strength should be increasing while by further increasing the coupling strength the population transfer vanishes and no transition would be occurred. Furthermore, it is noteworthy that by increasing the center frequency to 1, the FPG has an oscillating behavior. This oscillatory behavior in the non-adiabatic evolution is similar to the behavior of the population transfer in the non-adiabatic Landau-Zener approach [56]. This is while, the intermediate transitions under the non-adiabatic passage of the Landau-Zener method is also in accordance with our results [56, 57]. For center frequencies above 1, the FPG is independent of the center frequency value. Independence of the FPG to the center frequency for values greater than one can be explained by quadratic dependence of time on the laser frequency so that the center frequency no longer would be in resonance with the atomic levels. It is also obvious in this figure that the chirping parameter plays a prominent role in the enhancement of the system robustness. This means that the regions of complete transition are increased in such a way that the results do not change with small variations of parameters.

The FPG versus chirping parameter and coupling strength for different center frequency is depicted in Fig. 5. According to Eq. 5, this figure shows that the population transfer occurs in the non-adiabatic passage. The non-adiabatic transition probability has a sequence of maximum and minimum when the chirping parameter is swept. By increasing the chirping parameter, the complete transition regions are severely increased. This figure also confirms that robustness is considerably dependent on the chirping parameter such that by increasing the chirping parameter, robustness would be greatly enhanced as is seen in Fig. 4. It is also obvious in this figure that by increasing the center frequency, the robustness of the complete transition region would be increased. This effect of the center frequency and the chirping parameter on increasing of the robustness stems from the influence of these parameters on the evolution of avoided crossing areas. In the other words, reducing of regions of avoided crossing to one area increases the robustness (Fig. 3). Figure 5 also reveals that increasing the chirping parameters population transfer takes place in the higher coupling strength (Fig. 5(a)).

Fig. 6 shows the FPG versus chirping parameter and center frequency for different coupling strength. It reveals that by increasing the center frequency, the FPG has an oscillating behavior. In this figure, it is clear that by increasing the center frequency, blue strip width is increased so that the complete transition area is increased and the robustness reached the maximum. As mentioned before (Fig. 3(a, c)), this is related to the reduction of the avoided crossing regions. This is while the increase of the coupling strength has the same effect on the robustness. Furthermore, by comparing the different cases of Fig. 6 it is obvious that increase of the coupling strength causes the complete transition to be observed at the higher chirping parameter which it corroborates the results of Fig. 5. In addition, our findings confirm that the number of blue strips corresponds to the number of intermediate transition peaks.

The FPGs show that by appropriate adjusting of the laser chirping parameter, the center frequency and the coupling strength, complete and robust population transfer could be selectively achieved (Figs. 4,5 and 6) which the time of the complete transition could be drastically decreased compared to linearly traditional chirped laser (Figs.1 and 2).

## IV. CONCLUSION

In conclusion, because of the long adiabatic transition times in the traditional linear chirping method, quadratically chirped laser approach have been studied. In this method, in the non- adiabatic passage, robust population transfer has been studied when a quadratically chirped laser interacts with a two-level system. In the non-adiabatic passage model, the transition probabilities are also investigated towards the optimization of laser parameters for the robust transition control. In this case, by increasing the center frequency and the chirping parameter, FPG frequently changes. Also by increasing the coupling strength, the complete transition takes place for higher chirping parameters. It is also shown that increasing the chirping parameter and center frequency significantly increase the robustness of the control mechanism. In the non-adiabatic passage, while there are some unstable transitions at the time around zero, the final transition time drastically reduces. This work shows that the study of FPG according to various system parameters is a powerful method for robust engineering of the fast population transfer using quadratically chirped lasers in the nonadiabatic regime.

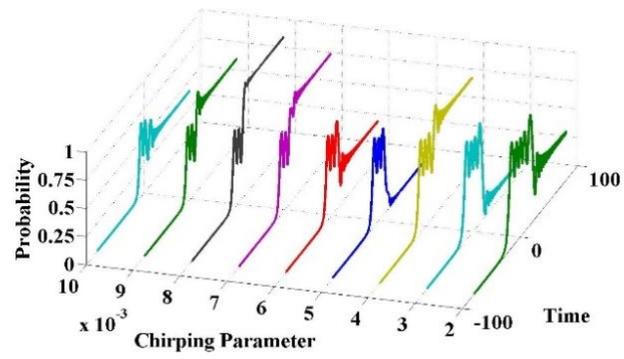

(a)

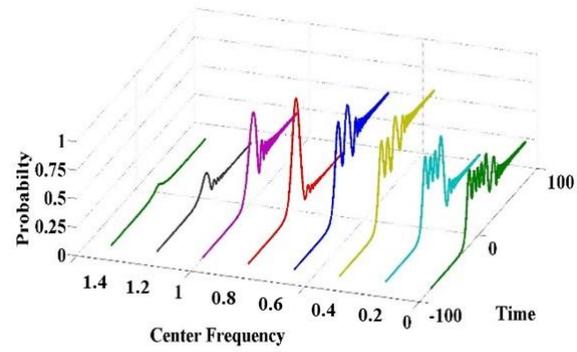

(b)

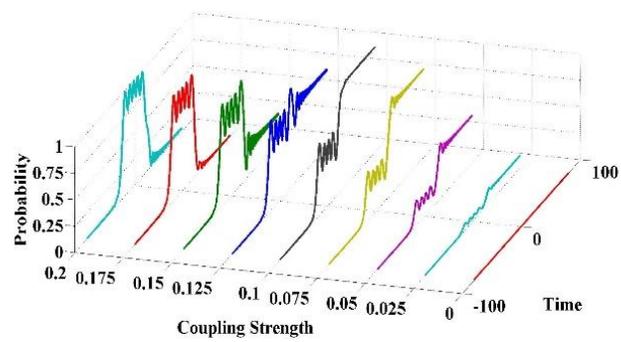

(c)

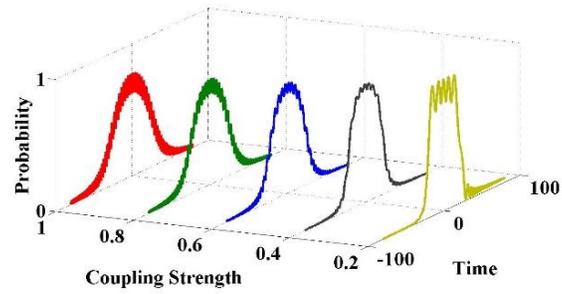

(d)

FIG. 1. Transition probability of quadratically chirped laser versus time and (a) chirping parameter for $\tilde{\omega}_{L0}$=0.1 and $\tilde{\omega}_I$=0.1 (b) center frequency for $\tilde{\alpha}$=0.002 and $\tilde{\omega}_I$=0.1 (c), (d) coupling strength for $\tilde{\alpha}$=0.002 and $\tilde{\omega}_{L0}$=0.1.

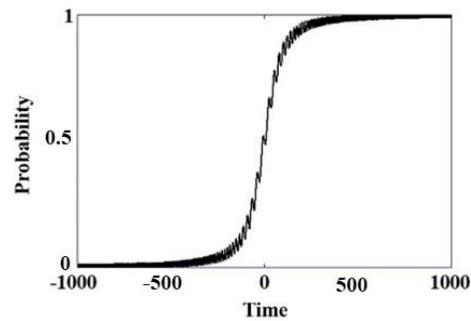

FIG. 2. Transition probability of linearly chirped laser versus time for $\tilde{\alpha}$=0.002, $\tilde{\omega}_I$=0.1 and $\tilde{\omega}_{L0}$=1.

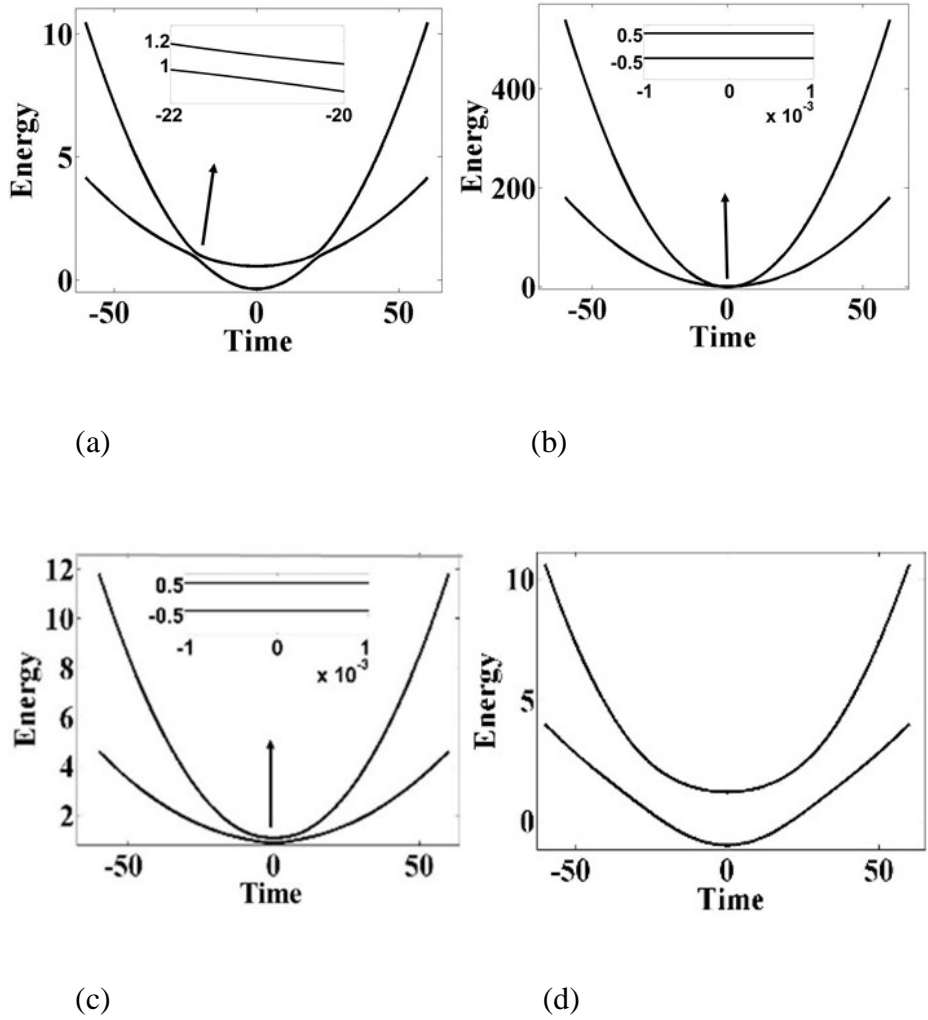

(a)

(b)

(c)

(d)

FIG. 3. Energy versus time for (a) $\tilde{\alpha}=0.002$, $\tilde{\omega}_1=0.1$, $\tilde{\omega}_{L0}=0.1$ (b) $\tilde{\alpha}=0.1$, $\tilde{\omega}_1=0.1$, $\tilde{\omega}_{L0}=0.1$ (c) $\tilde{\alpha}=0.002$, $\tilde{\omega}_1=0.1$ $\tilde{\omega}_{L0}=1$ (d) $\tilde{\alpha}=0.002$, $\tilde{\omega}_1=1$, $\tilde{\omega}_{L0}=0.1$.

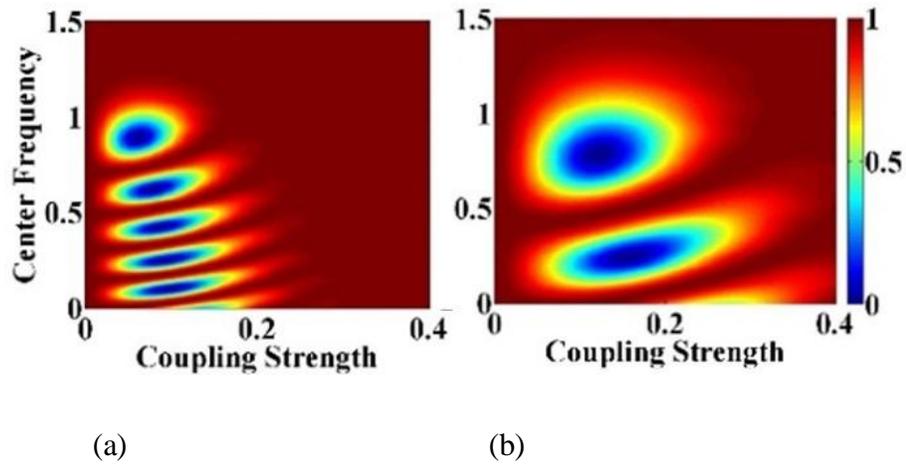

(a) (b)

FIG. 4. FPG versus center frequency and coupling strength for (a) $\tilde{\alpha}$=0.002 (b) $\tilde{\alpha}$=0.016.

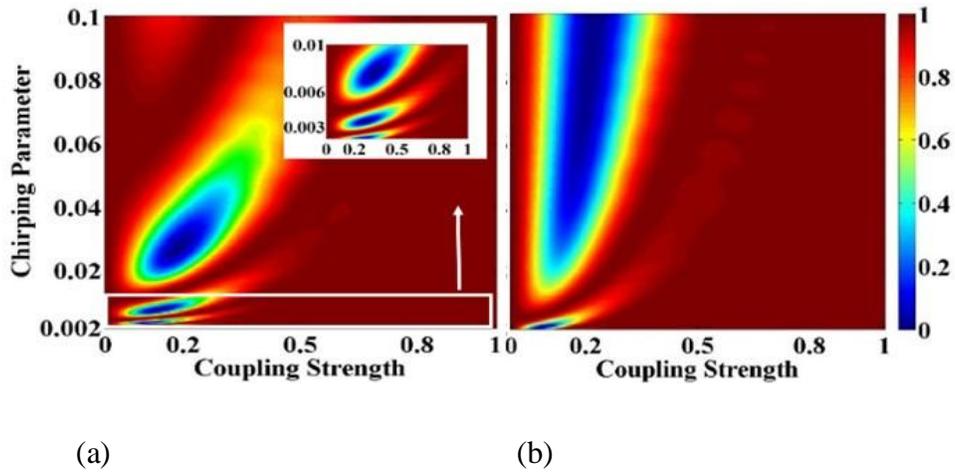

(a) (b)

FIG. 5. FPG versus chirping parameter and coupling strength for (a) $\tilde{\omega}_{L0}$=0.1 (b) $\tilde{\omega}_{L0}$=0.6.

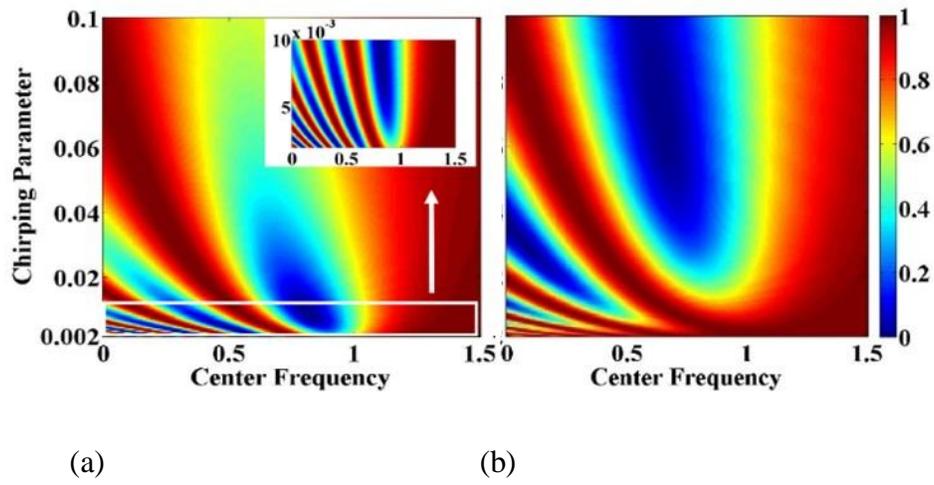

(a)                                    (b)

FIG. 6. FPG versus chirping parameter and center frequency for (a) $\tilde{\omega}_1$=0.05 (b) $\tilde{\omega}_1$=0.2